# On the coarsening mechanism and deformation of borides under annealing and creep in a polycrystalline superalloy

Lola Lilensten[1, #,*], Aleksander Kostka[2], Sylvie Lartigue-Korinek[3], Stoichko Antonov[4,5], Sammy Tin[4], Baptiste Gault[1,6], Paraskevas Kontis[1,*]

[1] Max-Planck-Institut für Eisenforschung, Max-Planck Straße 1, 40237 Düsseldorf, Germany

[2] Center for Interface-Dominated High Performance Materials (ZGH), Ruhr-Universität 44801, Bochum, Germany

[3] Institut de Chimie et des Matériaux Paris Est, UMR CNRS UPEC 7182, F-94320 Thiais, France

[4] Illinois Institute of Technology, 10 W. 32nd Street, Chicago, IL, 60616, USA

[5] Beijing Advanced Innovation Center for Materials Genome Engineering, State Key Laboratory for Advanced Metals and Materials, University of Science and Technology Beijing, Beijing 100083, China

[6] Department of Materials, Imperial College, South Kensington, London SW7 2AZ, UK

[#] Now at PSL Research University, Chimie ParisTech, Institut de Recherche de Chimie Paris, CNRS UMR 8247, Paris, France

[*] Corresponding authors: lola.lilensten@chimieparistech.psl.eu, p.kontis@mpie.de

**Abstract**

We have investigated the coarsening mechanism of intergranular Cr-rich $M_2B$ borides after creep and annealing at 850°C for approximately 3000 hours in a polycrystalline nickel-based superalloy. Borides were found to be coarser after creep, with measured thicknesses in the range of 800-1100nm, compared to borides annealed in the absence of external applied load (400-600nm). The borides had a thickness of 100-200nm before exposure at 850°C. Transmission electron microscopy revealed that coarsened borides have either the tetragonal

I4/mcm structure, or the orthorhombic Fddd, with those two structures coexisting in a single particle. The presence of a very high density of planar faults is systematically observed within the coarsened borides. The faults were correlated with chemical fluctuations of B and Cr, revealed by atom probe tomography. Our results allow us to suggest that borides coarsen by an epitaxy-like mechanism. In addition, partitioning of Ni and Co was observed at dislocations within the borides after creep providing insights into the deformation of borides. Consequences of coarsened intergranular borides on the creep performance of polycrystalline superalloys are discussed.

1. **Introduction**

Even as a minor elemental addition in both nickel- and cobalt-based superalloys, boron has major consequences on their microstructure and mechanical performance [1,2]. It is well documented that boron is typically found at grain boundaries, either segregated as an element in solid solution [3,4] or in the form of various borides, such as $M_2B$, $M_5B_3$ or $M_3B_2$ [2,5–10]. In the former case, it was suggested that interfacial boron increases the strength and cohesion of the grain boundary [11,12], thereby contributing to an overall ductility enhancement. However, in the latter case, borides have a contradicting character. In some cases, borides were found to be beneficial to the mechanical performance under creep conditions [13–15], whereas in other cases they act as dislocations sources that facilitate micro-crack initiation and micro-twinning under creep [1,16–18].

Many studies focused mainly on the composition of borides prior to deformation [5,9,19,20] or after creep of the alloy [15]. Yet, they do not provide insights into the microstructural evolution of intergranular borides, such as coarsening or dissolution, when subjected to thermal exposure with or without applied load. In addition, the deformation behavior of borides under creep at the near-atomic level has not yet been explored. Thus, a deeper understanding of the microstructural evolution and deformation behavior of the borides under creep is required to evaluate their role on the mechanical performance of polycrystalline superalloys.

Recent investigations detailing the deformation behavior of γ′ precipitates in Ni-based superalloys, have revealed interactions of solutes with crystal defects such as dislocations, stacking faults, microtwins or extended stacking faults within the γ′ precipitates leading to microstructural and chemical alterations [21–31]. In particular, it was shown that solutes, such as rhenium, chromium, molybdenum and cobalt are enriched at dislocations within the γ′ precipitates in nickel-based superalloys deformed at various temperatures [21,24,32–34]. This partitioning at crystal defects leads to microstructural and chemical alterations through what we referred to as the plasticity-assisted redistribution of interacting solutes (PARIS) mechanism in Ref. [35] that contributes to the deformation of γ′ precipitates.

However, similar studies on the interactions of solutes with crystal defects in other compounds, such as carbides, oxides and borides have not been considered so far. In this study, the polycrystalline nickel-based superalloy STAL15-CC was crept at 850°C under 185MPa load up to fracture (time to fracture approximately 3000 hours). We investigated the microstructural evolution of the intergranular Cr-rich $M_2B$ borides after deformation at 850°C. Scanning electron microscopy shows clear coarsening of the intergranular borides, while transmission electron microscopy reveals a high density of faulted planes within the coarsened borides. Atom probe tomography provides insights into the nature of solutes that segregate at crystal defects, faulted planes and dislocations, in the borides. In order to determine a possible role of the external applied stress on coarsening, samples of STAL15-CC were statically annealed at 850°C, i.e. in the absence of any externally applied load, for an equivalent time to the creep rupture time (i.e. 3000 hours). In this case borides were also observed to coarsen, and display a highly faulted structure.

Our findings allow us to investigate the coarsening behavior under thermal exposure at 850°C, as well as the deformation, with a focus on the interaction between solutes and defects occurring in these precipitates. This provides new insights into the microstructural evolution of borides and hence improves the current understanding of borides and their role on the mechanical performance of polycrystalline superalloys. Based on the mechanistic understanding we propose herein, accurate modelling and simulations can be designed, leading to a mathematical model of the microstructural evolution and influence on creep properties, which are beyond the scope of this particular study.

## 2. Materials and methods

The polycrystalline nickel-based STAL15-CC superalloy used in this study has a nominal composition of Ni-16.50Cr-5.50Co-0.60Mo-1.20W-10.00Al-2.40Ta-0.02Hf-0.5C-0.05B-0.01Zr (at.%). Following vacuum investment casting, specimens were hot isostatically pressed at 1150°C for 5 h under a pressure of 175 MPa to minimize internal casting porosity. This was followed by a solution heat-treatment at 1120°C for 4 h, air cooled, and further aging heat treated at 845°C for 24 h followed by air cooling. Creep specimens were machined from fully heat-treated bars, with a gauge length of 25mm and tested at 850°C under 185 MPa load up to fracture, the creep curve is provided in Figure S1 of the supplementary materials. Samples from fully heat-treated bars were annealed at 850°C for 3000 hours, and air-cooled. The annealing time was selected based on the corresponding creep rupture time for creep at 850°C and 185 MPa, which was approximately 3000 hours. Hereafter this sample is referred to simply as annealed.

The crept specimens were cut along the load axis and then polished with abrasive media down to 0.04µm colloidal silica surface finish. Same surface preparation was followed for the annealed samples. Backscattered electron (BSE) micrographs were recorded on a Zeiss Merlin scanning electron microscope (SEM) at an accelerating voltage of 20 kV and a probe current of 2 nA.

Specimens for transmission electron microscopy (TEM) were prepared from both crept and annealed samples using a FEI Helios Nanolab G4 CX focused ion beam system (FIB) operated at 30 kV. Low, 5 kV ion beam cleaning was applied for 2 minutes to each side of the TEM lamella in order to remove FIB beam damage. TEM imaging and diffraction were performed using a FEI Tecnai Supertwin F20 operated at 200 kV and equipped with a high angle annular dark-field (HAADF) detector. Atomic resolution image simulations were calculated using the MacTempas software [36].

Atom probe tomography (APT) specimens were prepared using a dual beam SEM/FIB instrument (FEI Helios Nanolab 600) via a site-specific lift-out protocol and mounted on a Si coupon [37]. APT measurements were carried out on a Cameca LEAP$^{TM}$ 5000 HR operated in laser pulsing mode at a pulse repetition rate of 125 kHz and a pulse energy of 55 pJ. The base temperature was set to 60K and the detection rate was maintained at 15 ions every 1000 pulses. Data analyses were performed using the IVAS 3.8.2 software package.

## 3. Results

### 3.1. Coarsening of borides under creep and annealing at 850°C

Figure 1 shows intergranular borides in the STAL15-CC alloy in different microstructural states. Figure 1a shows a grain boundary after full heat treatment. The grain size is approximately 750 µm, and a γ′ volume fraction of 51.8% is calculated in this alloy by the lever rule (see Figure S2 in the supplementary material) [38]. Although the lever rule applies in the case of dual-phase γ/γ′ microstructures, the amount of grain boundary precipitates in the present alloy is believed to be negligible and therefore not influence much the calculation of the γ′ volume fraction, especially for the heat-treated state considered for the calculation. The dark contrast Cr-rich borides can be more clearly seen in the close-up in Figure 1b, alongside MC carbides and large grain boundary γ′ precipitates [2,15]. As shown in Figure 1b the borides are rather small. Measurements performed on 145 borides in the heat-treated condition led to an average size of 210±80 nm. Results on this alloy's heat-treated state can be found in the literature [2,14,15].

Figure 1c shows a grain boundary and the borides after creep at 850°C, revealing microstructural alterations. In particular, the size of the borides has increased substantially, to 1060±340 nm (measurements performed on 74 borides). In order to separate the contributions of the deformation and that of the heat treatment seen by the sample during the creep test, a sample annealed at 850°C for 3000 hours was studied. The intergranular borides in that sample were also found to coarsen as in the case of the crept sample. Although the borides from the annealed sample are coarser compared to those in the heat-treated condition, with an average size of 720±290 nm (105 borides measured), they are still thinner compared to those after creep deformation. Note that Figures 1a, 1c and 1d are taken at the same magnification to make the extent of coarsening of the borides readily apparent, and that details of the size distribution are provided in supplementary materials as histograms.

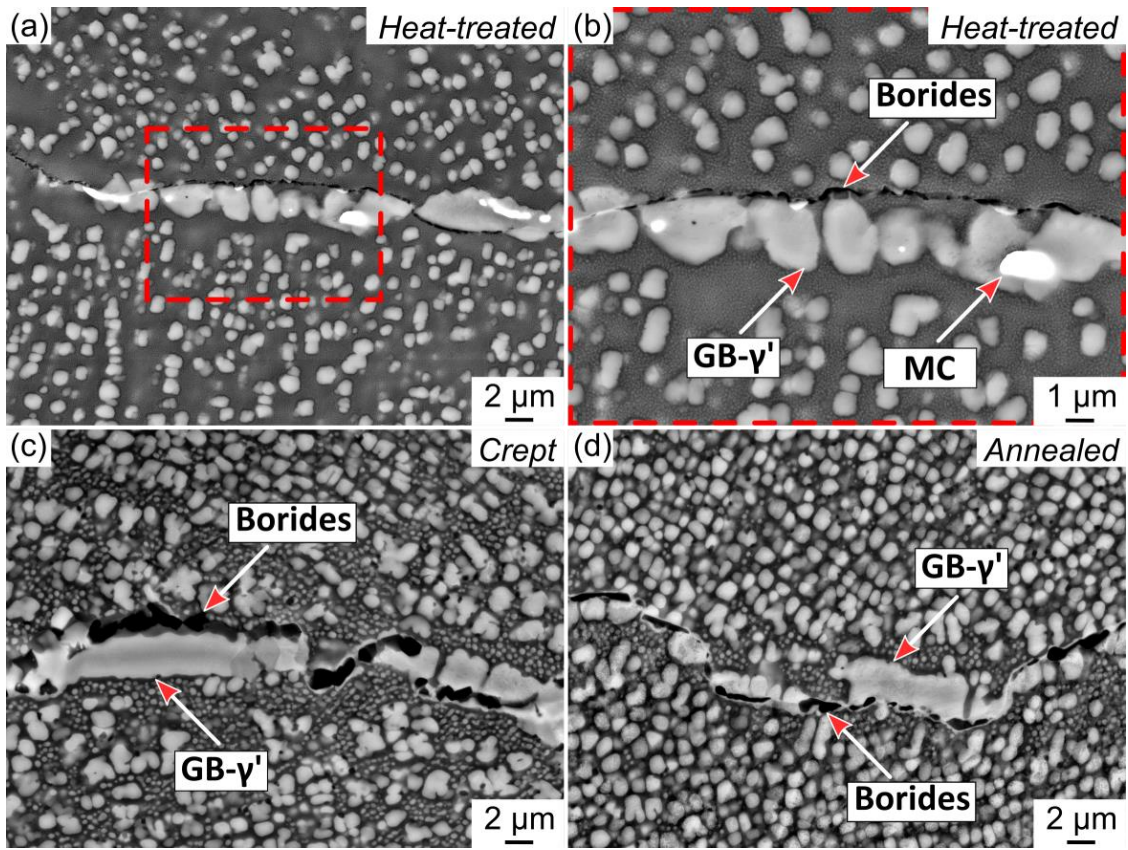

*Figure 1: Backscattered electron micrographs of STAL15-CC alloy: a-b) after full heat treatment, c) after creep deformation at 850°C under 185 MPa load up to fracture and d) after static annealing at 850°C for 3000 hours.*

**3.2. Structural information of coarsened borides at 850°C**

In order to understand the coarsening mechanism of the intergranular borides at elevated temperatures, TEM samples with coarsened borides were investigated for both crept and annealed conditions. Figure 2a shows a HAADF image from a grain boundary of the crept sample. Note that the TEM specimen was prepared parallel to the grain boundary. The matrix γ and γ′ precipitates are evidenced, along with several precipitates. Grain boundary borides, identified by their diffraction patterns, are adjacent to each other, and contain planar features. MC carbides are also identified (see green arrow in Figure 2a). Some of the large borides were characterized in more details. A TEM bright field image at higher magnification of an $M_2B$ boride, indicated by the red box in Figure 2a, is shown in Figure 2b. The diffraction pattern collected from this boride is shown as an inset in Figure 2b, and corresponds to a [111] zone axis of the tetragonal I4/mcm structure, with lattice parameters $a$=0.52 nm and $c$=0.43 nm, as reported in the literature [8,39,40]. One can note that this diffraction pattern is very close to that of a [001] zone axis of the Fddd orthorhombic structure of $M_2B$, as some $d_{hkl}$ distances are only 0.02 Å apart. Elongated streaks are visible, which indicate the signature of stacking faults in the {110} planes [41,42]. A high resolution TEM image of boride of Figure 2b is given in Figure 2c, evidencing more clearly the {110} faulted planes obtained after creep deformation at 850°C, in good agreement with the literature [8,39,40]. Finally, a dislocation (the image is rotated compared to Figure 2c to bring the (1-10) planes horizontal) is evidenced in Figure 2d, with an extended strain field. Its Burgers vector was partly determined following the Burgers circuit of Figure 2e, leading to a vector ±1/2[1-11], the slip plane being (10-1).

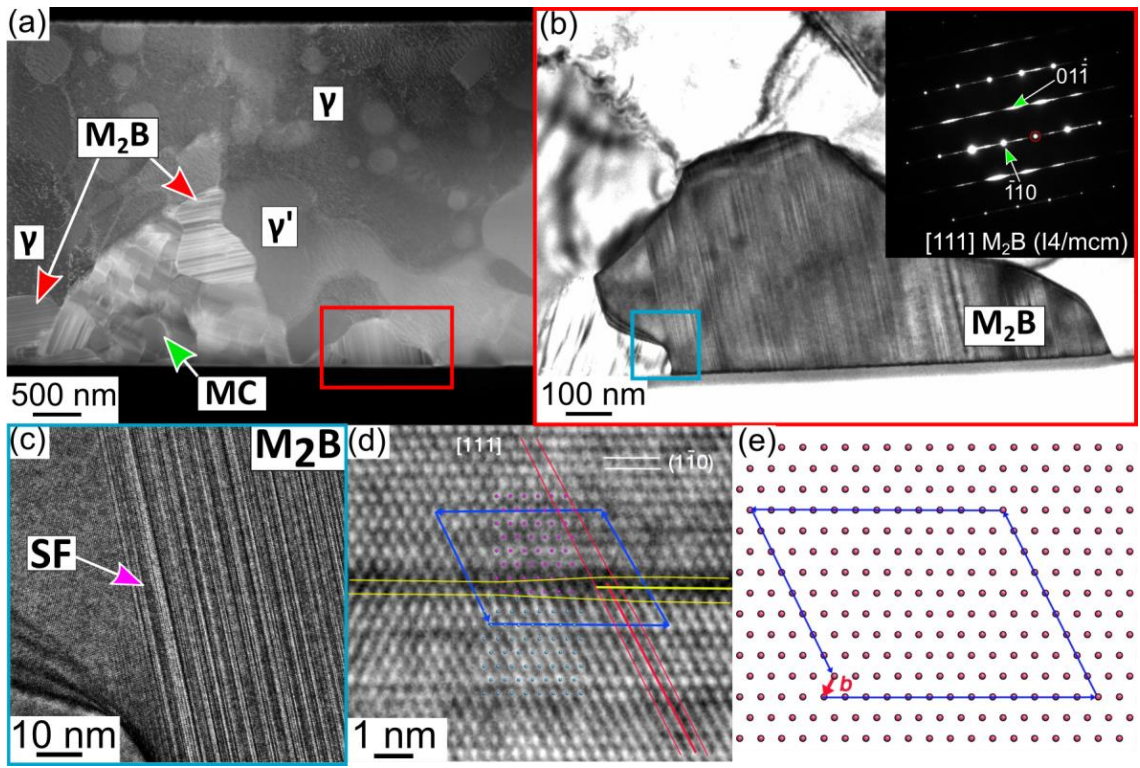

*Figure 2: TEM analysis of an intergranular $M_2B$ boride from the crept sample at 850°C. (a) TEM HAADF image of the sample showing several intergranular $M_2B$ borides (see red arrows for instance) and an MC carbide (green arrow) along the grain boundary. (b) TEM bright field image of the boride denoted by the blue box in Figure 2a alongside with its diffraction pattern. (c) High resolution TEM image showing details of the stacking faults in the boride from Figure 2b. (d) Image of a dislocation in the boride and (e) corresponding Burgers circuit.*

The investigation of another coarsened boride from the crept sample is summarized in Figure 3. Once again, the boride contains a high density of faulted planes. Its diffraction pattern can be indexed with the orthorhombic structure Fddd, with lattice parameters $a$=0.42 nm, $b$=0.73 nm and $c$=1.46 nm. The diffraction pattern given in the inset of Figure 3a shows streaks suggesting that the stacking faults are in the {002} planes. The high resolution TEM image in Figure 3b along with the provided modelling of the Fddd structure shows indeed the stacking of {200} planes, with a fault. However, some of the planar features observed in the boride of Figure 3a may also correspond to a change in the crystallographic structure, as shown in Figure 3c. The high-resolution TEM image of this area, from the boride shown in Figure 3a, clearly shows the two structures, Fddd and I4/mcm, successively stacked on top of each other, with the relationship [110]Fddd // [001]I4/mcm.

Overall, the faulted planes in the borides can be identified as normal stacking faults, twins (not detailed here, for more information see [8]) or a change of the crystallographic structure, with all these possibilities coexisting in a single precipitate.

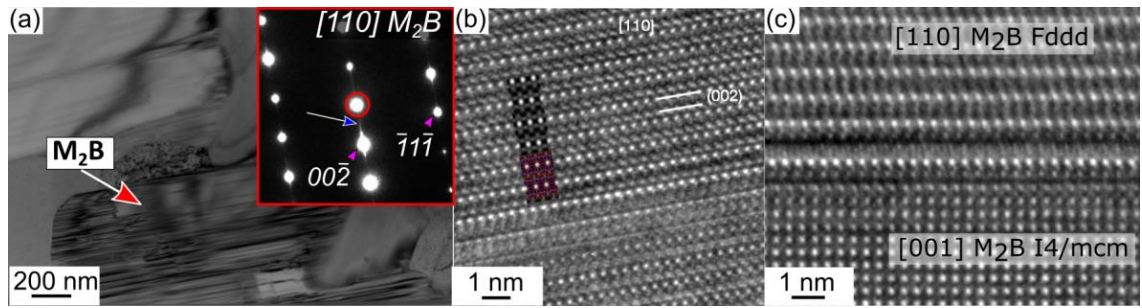

*Figure 3: TEM analysis of an intergranular $M_2B$ boride from the crept sample at 850°C. (a) TEM bright field image of the boride along with its diffraction pattern. (b) High resolution TEM image of the boride in (a) showing details of the stacking faults in the boride, along with image simulation. In the superimposed structure, Cr atoms are in pink and B atoms in orange. The white dots correspond to B columns (defocus -56 nm, thickness 20nm). (c) High resolution TEM image of the boride in (a) showing two M2B structures, Fddd and I4/mcm stacked on top of each other.*

In the case of the annealed sample at 850°C for 3000 hours (without applying any external stress), similar observations were performed within the coarsened borides. Figure 4a is an HAADF image of a TEM sample taken along a grain boundary containing coarsened $M_2B$ borides. Figure 4b clearly shows that coarsened borides within the annealed sample are heavily faulted and these stacking faults within the $M_2B$ borides are not the result of deformation under external load. A diffraction pattern taken from the boride in Figure 4b allows to characterize a $M_2B$ boride exhibiting the Fddd crystal structure in [110] zone axis, and the streaks in the diffraction pattern (in the same zone axis than the boride of Figure 3) also characterize faults in the {002} planes, evidenced in Figure 4c. Some dislocations are also visible inside the matrix, in the vicinity of the coarsened boride (see green triangles in Figure 3b).

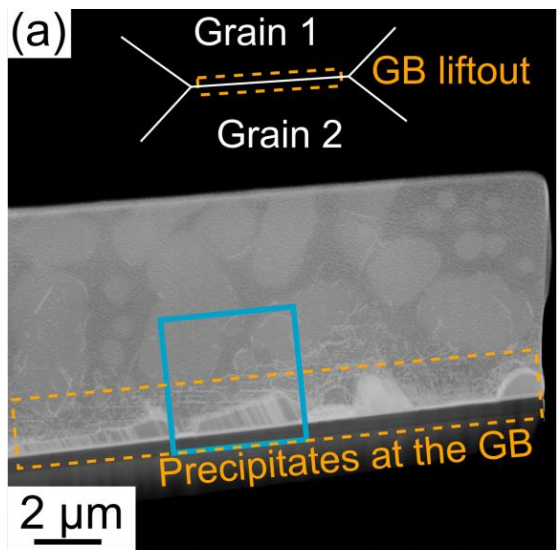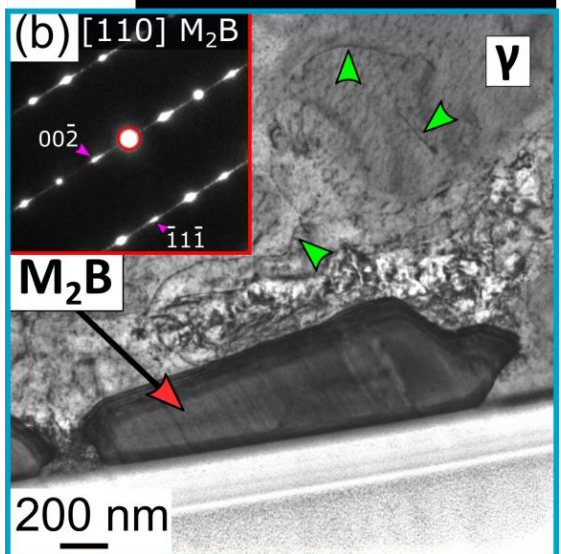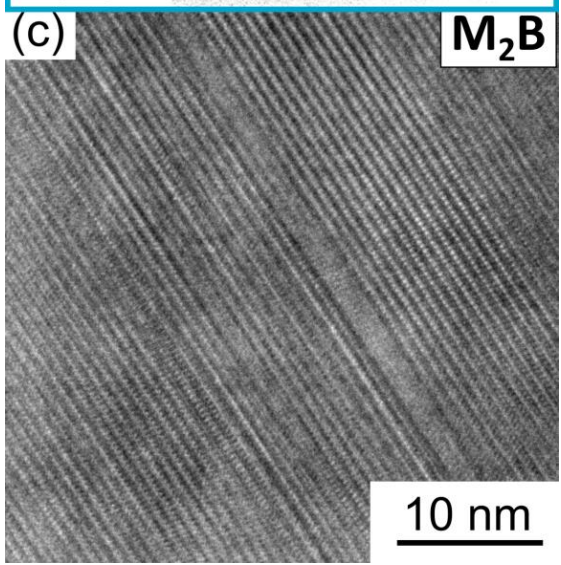

*Figure 4:* *TEM analysis of an intergranular* $M_2B$ *boride from the sample annealed for 3000 hrs at 850°C without load. (a) Schematic of the GB lift-out and TEM HAADF image of the lamella, showing precipitates along the grain boundary (orange dashed rectangle). (b) TEM HAADF image of the area denoted by the blue box in Figure 4a, showing a higher magnification of a boride, and the corresponding diffraction pattern. Green arrows point at dislocations in the neighboring γ matrix. (c) High resolution TEM image of the faulted planes of the borides in (b).*

### 3.3. Interactions of solutes with crystal defects in borides

APT specimens were prepared from both crept and annealed specimens, aiming to investigate interactions of solutes with the stacking faults taking place during the coarsening of the borides.

Figure 5a shows an APT reconstruction from a coarsened intergranular boride from the crept sample at 850°C. A 2D density map of Cr in Figure 5b, corresponding to the area denoted by the rectangle in Figure 5a, reveals the presence of compositional variations along planes that cross the entire boride. These variations are likely associated with the presence of the planar faults observed by TEM. Due to the angle between the planes and the needle-shaped specimen's evaporation axis Z, the possibility that these planes correspond to a field evaporation artefact can be disregarded.

A 1D composition profile was extracted from an elliptic cylindrical region of interest taken along the arrow #1 in Figure 5b, is shown in Figure 5c. Composition fluctuation of Cr and B can be seen, where Cr decreases in regions where B increases. In particular, five B-rich regions were identified, with B enrichment varying from region to region between 25.0 and 28.0 at.%. The lower B-content characterized here compared to the expected stoichiometry of $M_2B$ can be explained by the detector dead-time, and was already reported for boron [43], similarly to results on nitrogen, oxygen and carbon [44,45]. Besides the enrichment of B on planar features, segregation at dislocations was also observed: tubular features, which correspond to elemental segregation at dislocations, also appear within the APT reconstruction in Figure 5a. Figure 5d shows a 1D composition profile across a dislocation. The 1D composition profile was extracted from a cylindrical region of interest perpendicular to the dislocation indicated by the arrow #2 in Figure 5a. An enrichment of Ni and Co at the dislocation core up to approximately 6.1 and 2.4 at.% is revealed, respectively. A concurrent depletion of B and Cr was also observed as shown in Figure 5d.

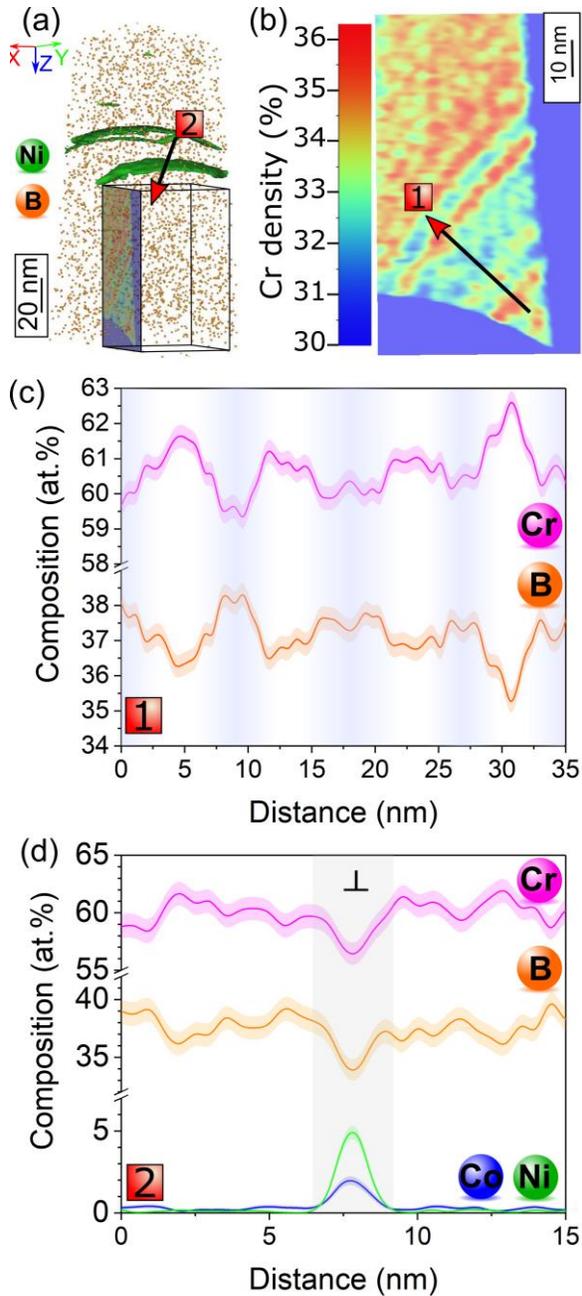

*Figure 5: APT analysis of an intergranular $M_2B$ boride from the crept sample at 850°C. (a) Atom probe reconstruction from the boride containing dislocations highlighted by iso-composition surfaces of 0.8 at.% Ni. (b) 2D density map for Cr from the box in Figure 5a. (c) 1D concentration profile across the B-rich planes indicated by the arrow #1 in Figure 5b. The B-rich planes are highlighted by a blue background area. (d) 1D composition profile across the dislocation in Figure 5a, along the arrow #2, showing an increase of Ni and Co at the dislocation core. Error bars are shown as lines filled with color and correspond to the 2σ counting error.*

In order to unambiguously reveal the interactions of solutes with the stacking faults and dislocation in borides, we have collected additional APT data. Figure 6a shows a second APT 3D-reconstruction from a coarsened intergranular boride from the crept sample.

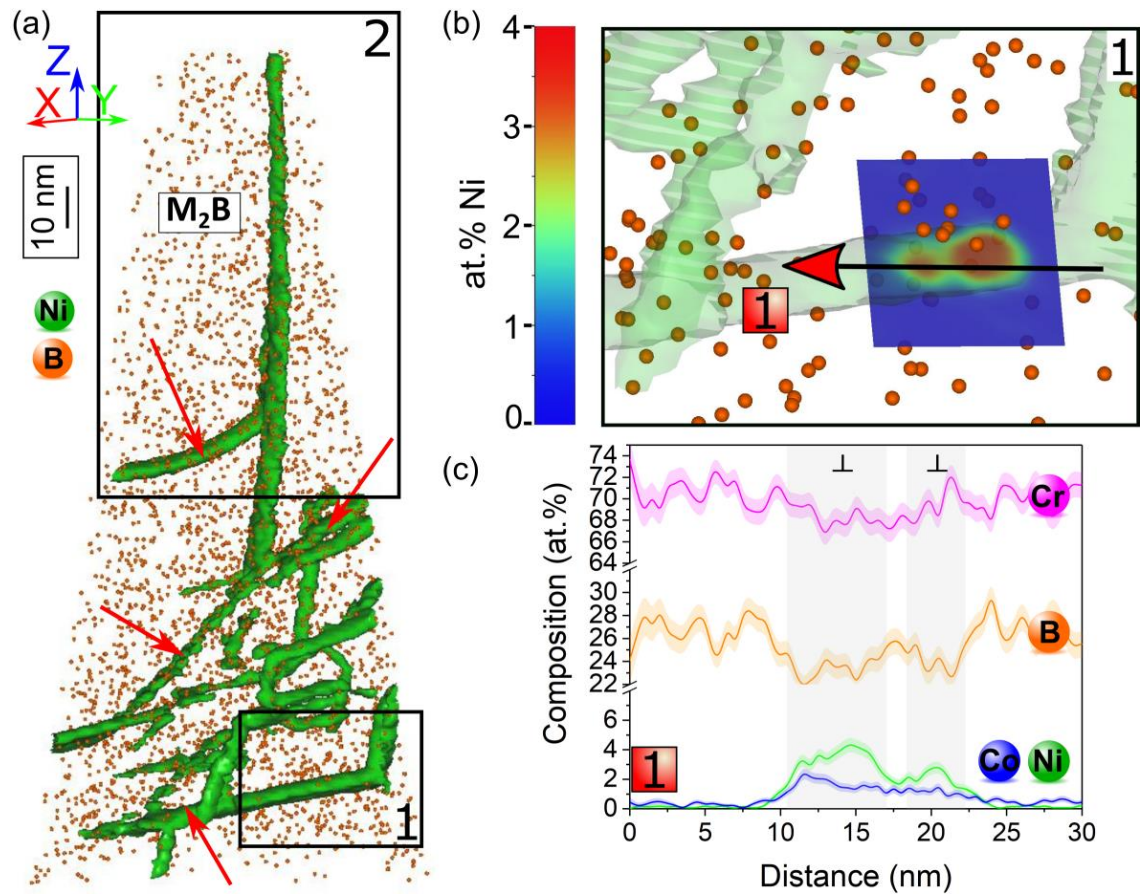

*Figure 6: APT analysis of an intergranular $M_2B$ boride from the crept sample at 850°C. (a) Atom probe reconstruction from the boride. Dislocations, highlighted by iso-composition surfaces of 1.0 at.% Ni., are denoted by red arrows. (b) Magnified view of the APT reconstruction from the rectangle 1 in Figure 6a, with a 2D concentration map plotted across a dislocation for Ni. (c) 1D concentration profiles across the dislocation indicated by the red arrow in Figure 6b. Error bars are shown as lines filled with color and correspond to the 2σ counting error.*

Similar Ni- and Co-rich dislocations as those reported in Figure 5 are observed. In this case, the dislocations appear to be in two different crystallographic planes, however this needs to be confirmed by further TEM investigations. Figure 6b shows a close-up on the region-of-interest indicated by the rectangle #1 in Figure 6a. A 2D composition map is plotted for Ni across the dislocation revealing two maxima and suggesting that this dislocation is a pair of dislocations within the boride. A composition profile is extracted from an elliptical cylinder region of interest perpendicular to the dislocation, along the red arrow #1 (Figure 6b), and is displayed in Figure 6c. It confirms that two distinct dislocations are observed, enriched in Ni and Co, as for the dislocation in Figure 5, with contents up to 4.4 at.% and 2.6 at.%, respectively. Boron and chromium seem again to be depleted at the dislocation core.

B-rich planes are also evidenced in this dataset, as indicated in the 2D compositional map provided in Figure 7a, corresponding to the box denoted as #2 in Figure 6a. A 1D composition profile perpendicular to these planes (arrow #1 in Figure 6a) is plotted in Figure 7b for B and Cr, and in Figure 7c for W and Mo. It can be seen that Cr and B fluctuate with opposite trends, as already shown in Figure 5c. Mostly, a large decrease in the Cr composition from 70 down to 68.5 at.% is visualized at about 9 nm, accompanied by a B enrichment up to 28.1 at.% (from about 26 at.%). Besides Cr and B fluctuations across the stacking faults, compositional variations of W and Mo were also observed as shown in Figure 7c. Although the B-rich planes do not clearly overlap with variations of W and Mo, as highlighted by the absence of relationship between the W/Mo composition profiles and the blue areas in Figure 7c corresponding to the B-rich planes, such fluctuations within a heavily faulted microstructure need to be considered. Even more clear fluctuations of W and Mo are shown below in a different sample and will be discussed in section 4.1 in detail.

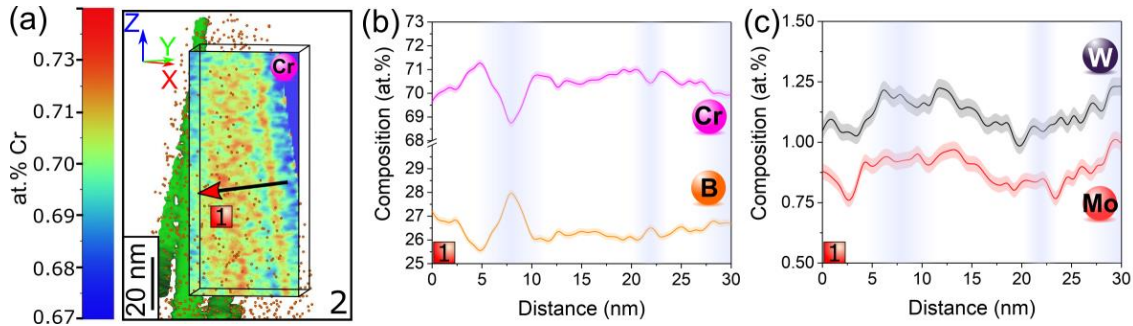

*Figure 7: (a) Magnified view of the APT reconstruction from the rectangle #2 in Figure 6a, superimposed with a 2D concentration map for Cr. (b) and (c) 1D concentration profiles across the B-rich planes indicated by the arrow #1 in Figure 7a for B and Cr and for W and Mo, respectively. The B-rich planes are highlighted by a blue background area. Error bars are shown as lines filled with colour and correspond to the 2σ counting error.*

APT analysis was also performed for intergranular coarsened borides from the annealed sample at 850°C containing stacking faults. Figure 8a shows an APT reconstruction containing a boride and its interface with the γ matrix. Planar fluctuations of Cr are observed as shown in the 2D compositional map in Figure 8b, extracted from the region indicated by the black dashed box in Figure 8a. A 1D-composition profile across the planar compositional variations as denoted by the arrow #1 in Figure 8a, is plotted in Figure 8c. As in the crept sample, B-rich planes (highlighted by a blue background) are also observed within the boride of the annealed sample, with Cr fluctuations being the opposite than B fluctuations.

In addition, the composition profile for W and Mo across the B-rich planes is given in Figure 8d, and evidences an increase in W from 1.2 to 1.6 at.% and in Mo from 0.5 to 0.8 at.%. Besides, similarly to Figure 7c, Mo and W display similar variations, meaning that both increase or decrease at the same time. The blue background areas corresponding to the Cr and B fluctuations are provided in Figure 8d for comparison with the fluctuations of W and Mo, and show that the fluctuations of B/Cr and W/Mo do not coincide. Species-specific field evaporation conditions could lead to variations in the spatial resolutions and an associated blurring of the compositional peaks [46], however, the low composition of these species in this case makes this possibility rather unlikely. Thus, in the case of W and Mo, compositional fluctuations do not appear correlated to B/Cr planar features.

Finally, in Figure 8e, a composition profile as a function of the distance to an iso-composition surface with a threshold of 10 at.% Ni (i.e. proximity histogram or proxigram) was plotted across the interface between $M_2B$ and γ, see arrow #2 in Figure 8a [47]. It clearly shows that the composition in W and in Mo is higher in the matrix than in the boride, and that an enrichment of Mo is observed at the interface.

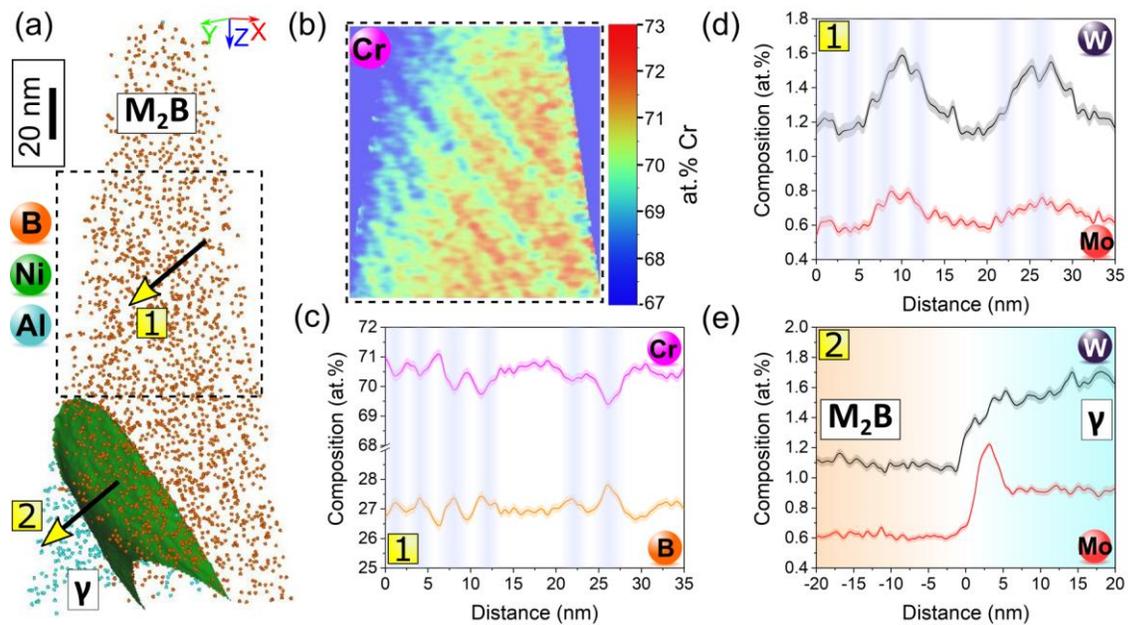

*Figure 8: APT analysis of an intergranular $M_2B$ boride from the annealed sample at 850°C. (a) Atom probe reconstruction from the boride and its interface with the γ matrix. The interface is shown with an iso-composition surface at 10 at.% Ni. (b) 2D concentration map for Cr corresponding to the black dashed boxed region in 8a. (c) and (d) 1D concentration profiles along the yellow arrow #1 in 8a for W and Mo and for Cr and B respectively. (e) Proxigram showing the W and Mo composition at the $M_2B$ and γ interface (arrow #2 in Figure 8a). Error bars are shown as lines filled with colour and correspond to the 2σ counting error.*

## 4. Discussion

The present study investigates the coarsening behavior of intergranular borides in a polycrystalline Ni-based superalloy during creep deformation at 850°C, using primarily TEM and APT. In order to separate the effects of temperature and load, samples annealed at 850°C without any external applied load and for a similar duration were also investigated. Based on the experimental observations, the discussion will be first focused on the coarsening mechanism of the borides. Their deformation will be treated next. Finally, the consequence of the coarsening of borides on the mechanical properties will be discussed.

### 4.1. Coarsening mechanism of borides at elevated temperatures

Comparison of the borides present in the initial heat-treated microstructure to those observed after creep or in the annealed sample in Figure 1 reveals that extensive coarsening of the borides occurred. APT analyses evidenced planar compositional fluctuations for W and Mo in both the crept and the annealed samples. In the analyzed borides, the average W and Mo compositions are about 1.0±0.04 at.% and 0.6±0.04 at.%, respectively. Within the planar fluctuations, W and Mo concentrations are measured to increase up to approximately 1.6 at.% and 0.8 at.% (Figure 7c and 8d), respectively. Interestingly, the compositions of W and Mo vary in a synchronous manner as they either both increase or decrease together. These fluctuations, as well as the proxigram given in Figure 8e, suggest that W and Mo are segregated away from the coarsened boride.

The compositional variations could thus be attributed to the growth of the boride, and are illustrated by a schematic in Figure 9 for W, considering that a similar behavior is expected in the case of Mo. As the boride grows, W is segregated away and diffuses outwards, towards the matrix (Figure 9a). Yet, considering the slow diffusion of W, it is possible that the rate of boride coarsening is faster than the diffusion rate of W. Therefore, repelled W piles up at the interface between the boride and the matrix (Figure 9b), up to a point when the coarsening is faster than the W diffusion. As a consequence, some amount of W is integrated in the newly grown boride and forms layers (Figure 9c), as shown by APT in Figures 7c and 8d.

Finally, as the boride keeps coarsening, the pattern repeats itself (Figure 9d and 9e). This suggests that the W/Mo dense planes are perpendicular to the growth direction, which matches the APT results of Figure 8a, showing W/Mo planes observed parallel to the M$_2$B/γ interface.

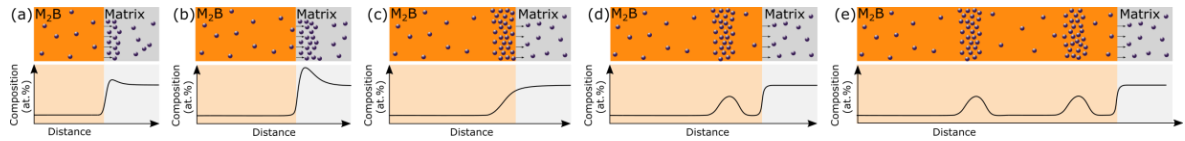

*Figure 9:* Schematic illustration of the proposed sequence of the interaction between W/Mo (represented as black circles), the boride and the matrix, during the coarsening of M$_2$B boride.

Then, one can note that W/Mo-rich planes are also parallel to the characterized B-rich planes evidenced by APT analyses, which seem to extend across the analyzed APT volume. Cr typically fluctuates between 70 and 71 ± 0.17 at.%, with maxima at 72 at.% and minima at 69 at.%, and B fluctuates between 26 and 27.5 ± 0.17 at.%, with minima at 25.5 at% and maxima at 28 at.%. Besides, the Cr fluctuations show an opposite trend to the B fluctuations. The fluctuations of W/Mo on one hand and Cr/B on the other hand, although observed in parallel planes, do not happen at the same place, suggesting that they are not correlated.

The numerous B-rich planes are likely to be correlated to the high density of faulted planes in the coarsened M$_2$B borides. Although the formation of deformation induced stacking faults in MB-type borides has been previously reported [48], based on the present results, it seems that the present planar faults are not mechanically induced, since they are also observed in the case of the annealed sample, with no external load.

The coarsened borides exhibit a tetragonal structure with space group I4/mcm or an orthorhombic one with space group Fddd. They are always heavily faulted, with stacking faults in their {110} planes (tetragonal structure) or in their {002} planes (orthorhombic structure). Such faults have been reported before [8,9,39,40], and some can be related to the intergrowth of the two structures, tetragonal and orthorhombic, for which two possible orientation relationships have been reported: [001]$_{tetragonal}$//[100]$_{orthorhombic}$ and $(110)_{tetragonal}//(001)_{orthorhombic}$ (or normal intergrowth) on one side, and $[001]_{tetragonal}//[110]_{orthorhombic}$ and $(110)_{tetragonal}//(001)_{orthorhombic}$ (or twin-mediated intergrowth) on the other side [8,40]. The case of intergrowth of these two structures is indeed observed in the present case (figure 3). As mentioned by Goldfarb et al. [39], the two structures are actually very close, since only a slight adjustment of B atoms lying on the {110} atomic planes is needed to transform the $Cr_2B$ structure from I4/mcm to Fddd. Besides, they share the same building polyhedron, as evidenced by Hu et al. [8].

A hypothesis would thus be that the B fluctuations are correlated to the planar faults or structural changes in the $\{110\}_{I4/mcm}$ planes or the $\{002\}_{Fddd}$ planes characterized by TEM. Considering that B-rich planes are parallel to W/Mo rich planes, it could be suggested that the borides coarsen along the <110> direction (tetragonal structure) or along the <001> direction ($M_2B$ Fddd), following an epitaxy-like mechanism of $\{110\}_{I4/mcm}$ and $\{002\}_{Fddd}$ planes stacking on top of each other. The high density of faults would then be a consequence of growth defects, with mistakes in the stacking sequence that lead to the introduction of a fault or of another structure, that extend across the whole precipitate.

The intergrowth of tetragonal and orthorhombic $M_2B$ is expected to be neutral from a composition point of view. Published electron energy loss spectroscopy (EELS) measurements reported a homogeneous distribution of B and Cr across the faults [40], but the length over which the composition was measured by EELS is very large (600 nm), and this would not have been able to resolve the small, local variations that APT was able to unveil in this study, in the range of about ten nm and of a few at.%. A hypothesis to explain the compositional changes, becoming off-stoichiometry, could be that the stacking of the building polyhedron leads to the local formation to another phase with different stoichiometry, such as $M_5B_3$ or $M_3B_2$. Intergrowth of these last two structures has been reported before, and the $M_5B_3$ structure contains building polyhedron of $M_2B$ [49]. Moreover, layers of $M_2B$ in $M_5B_3$ have been observed [49]. However, such intergrowth occurs in the tetragonal <001> direction, which is not what is observed here. Besides, HRTEM could not evidence the presence of $M_5B_3$ or $M_3B_2$ in the present case. Another possibility would be that there is occasionally a loss of symmetry of the Fddd structure, that would accommodate non-stoichiometric compositions, as observed in $M_2B$ borides of Co-Re-Cr superalloys [50]. Reaching locally such non-stoichiometric states, possibly influencing the crystal structure, may be due to different diffusion rates between Cr and B, leading to lesser available amounts of one of the solutes with respect to the other. However, in order to fully interpret the complex compositional fluctuations of B and Cr that are reported here for the first time, more systematic structural and compositional investigations are required, particularly in the early stages of coarsening of the borides.

With the hypothesis that the B/Cr fluctuations are related to a possible structural change, the W and Mo solubility could be higher in one of the structure/phases, leading to a better integration in one of the structures, and to the W/Mo compositional variations. Similarly, increased composition of W/Mo at the interface could lead to the formation of another structure, that would be more tolerant to W and Mo. However, in such cases, fluctuations would correlate between the B-rich planes and the W/Mo variations, which is not the case. Therefore, it is believed that W and Mo fluctuations are only related to the diffusion of these solutes from the boride towards the matrix and not to the crystal structure.

Overall, based on the experimental APT results in this study, a boride coarsening process that occurs via epitaxial layer-by-layer mechanism is proposed. Crystallographic stacking is expected to happen in the $\{110\}_{I4/mcm}$ and $\{002\}_{Fddd}$ planes. This stacking produces planar faults that extend through the whole boride, as well as enables intergrowth of the two structures. Such faults are believed to be related to planar compositional variations in B and Cr. During the growth, W and Mo are also repelled out of the borides, but due to a low diffusion rate, with small amounts incorporated in the structure, following a pattern described in Figure 9. Our SEM observations clearly shows that the coarsening of the boride is enhanced by the externally applied stress, as the borides are larger in the crept sample than in the annealed one. A possible explanation could be related to the higher amounts of Cr at grain boundary regions brought by the dislocations following the pipe diffusion mechanism [25]. However, further investigation is required in order to understand the coarsening rate of the borides, and to fully relate the B/Cr fluctuations to the faulted planes.

**4.2. Insights on the deformation of the borides under creep**

An evidence of the presence of dislocations in a boride after creep is given in Figure 2g and 2h. The APT analyses also evidenced the presence of dislocations in the crept samples. The results presented here come from two different datasets obtained from samples from different locations, allowing us to suggest that it is not an isolated observation. In both cases, dislocations seem to follow two possible directions, with one single dislocation that may consists of several segments pertaining to the two directions. As depicted in Figures 5d and 6c, some dislocations seem to go by pairs, whereas some others are evidenced as single dislocations.

Imaging dislocations in the APT datasets is enabled by the partitioning of Ni and Co at the dislocations, with compositions as high as approximately 4 at.% and 2 at.%, respectively, which is substantial considering their base compositions are below 0.2 at.% for Ni and 0.5 at.% for Co. An extended core, as the one characterized by TEM in Figures 2g and 2h, could also accommodate large amounts of solutes. Segregation of solutes at defects, such as dislocations, is driven by the minimization of the system's free energy. This can lead to either conventional Cottrell- or Suzuki-type decoration or even to spatially confined linear chemical-structural transformation states [51,52]. Both effects can be seen as akin to a partitioning phenomenon between the boride matrix and the defect and hence depend on the composition of the borides as well as of the surrounding phases in contact with the dislocations [21]. In both cases reported here, Ni and Co partition to the dislocations, while the concentrations of those two solutes within the borides are low, especially compared to their concentration in the $\gamma$ matrix (see [15]). Therefore, it is suggested that the observed solutes segregated at the dislocations' cores have most likely moved along the dislocations line connected to the matrix via pipe diffusion rather than having been collected during the dislocation glide inside the boride.

One could argue that some of the observed dislocations are related to the misfit between the structures involved in the coarsening. However, no dislocations were observed by APT in the annealed sample to confirm such a scenario, and stacking of tetragonal and orthorhombic $M_2B$ has been reported to lead to coherent interfaces, without lattice misfit dislocations [8].

Therefore, the present results rather suggest that the observed features are deformation-induced dislocations. Significant plasticity in borides deformed in the same temperature range has also been observed and interpreted by the nature of chemical bonding [53].

### 4.3. Impact of borides coarsening and deformation on the superalloy properties

Based on the results, borides seem to be able to deform plastically, and their dislocations cores are enriched in Ni and Co. This suggests that the local stresses built up on the interfaces during creep can be released by the plastic deformation of boride particles, which could improve the creep resistance. It also suggests that chemical interactions with the surrounding matrix occur, by leaching Ni and Co. However, high-temperature annealing and, to a further extent, high-temperature creep, lead to coarsening of the boride, affecting the microstructural stability of the grain boundaries during creep. The coarsening rate may be the thus be a life limiting parameter for the superalloy. Agglomeration of borides at grain boundaries forming a continuous film can potentially act as crack initiation sites, particularly at grain boundaries transverse to the applied stress during creep. This particular alloy has indeed been shown to be very sensitive to grain boundary cracking, and so the coarsening of borides may limit the creep properties of the alloy, in spite of the $\gamma/\gamma'$ microstructure, with conventional $\gamma'$ volume fraction, displaying traditionally good creep resistance [2,14,15,54]. This is similar to what was shown before in the case of intergranular $M_{23}C_6$ carbides in superalloys [55]. The continuous coarsening will leach elements from the surrounding matrix such as Cr, but also Ni and Co that diffuse towards the dislocations cores, leading to chemical and microstructural destabilization, and changes in $\gamma/\gamma'$ equilibrium. It has been previously shown that in the case of STAL15-CC alloy at the vicinity of coarsened borides there is a region depleted of $\gamma'$ precipitates [15]. Hence, the combination of a continuous film of borides at grain boundaries with regions depleted of $\gamma'$ precipitates can potentially facilitate the formation of micro cracks. It becomes apparent that precipitation of secondary grain boundary particles, their coarsening rate and their deformation requires careful attention in order to design advanced polycrystalline superalloys.

## 5. Conclusions

The behavior of intergranular Cr-rich $M_2B$ borides under creep and annealing was studied for the polycrystalline nickel-based superalloy STAL15-CC, in order to better understand the behavior of these precipitates, and how they affect the overall properties of the material. The following conclusions can be drawn:

- Borides were found to coarsen during creep deformation at 850°C after approximately 3000 h and during static annealing at 850°C for 3000 h. Borides are coarser after creep compared to after annealing.
- $M_2B$ borides are characterized, with tetragonal I4/mcm or orthorhombic Fddd structure. The coarsened borides display a high density of stacking faults in the {110} planes (tetragonal structure) or in the {002} planes (orthorhombic). Intergrowth of both structures is also reported.
- Planar fluctuations of W and Mo are observed within the borides. Due to a lower solubility in the borides, these elements are expected to be repelled out as the borides coarsen. However, due to their slow diffusion compared to the high coarsening rate of the boride, some of the W/Mo that could not diffuse away still get incorporated.
- Planar fluctuations of B and Cr were characterized by APT. It is proposed that they are related to the observed planar faults or to the structural changes due to the intergrowth, which led us to propose a coarsening mechanism for borides as follow: coarsening proceeds following an epitaxy-like mechanism, with stacking in the $\{110\}_{I4/mcm}$ and $\{002\}_{Fddd}$ planes, leading to a very complex structure made of several faults and intergrown tetragonal ad orthorhombic $M_2B$.
- Borides display plastic deformation and dislocations are evidenced by TEM. Enrichment of Ni and Co solutes up to 6.1 and 2.4 at.%, respectively, was measured by APT at the dislocations within the borides of the crept sample. The solute segregation at the dislocation is believed to be pumped out of the surrounding matrix, leading to an alteration of the latter's composition.

The present findings emphasize that borides take an active place in the evolution of superalloys during annealing or creep. Their coarsening behavior is rationalized by a layer-by-layer mechanism. Their growth and the first insights on their plasticity evolve solute transfer, of Cr that is the main component of the borides, but also of Ni and Co that appear to be incorporated in the borides, as Cottrell atmospheres at dislocations. Such effects will likely have consequences on the overall evolution of the grain boundary structure during creep, and a better understanding of these local mechanisms will allow to improve design strategies of such safety critical materials.

**Acknowledgements**

P.K. thanks Siemens Industrial Turbomachinery for provision of the material and performing the creep and annealing tests. Uwe Tezins & Andreas Sturm for their support to the FIB & APT facilities at MPIE. L.L. thanks P. Vermaut for fruitful discussions. We are grateful for the financial support from the Max-Planck Gesellschaft via the Laplace project for both equipment and personnel (P.K.). P.K. also acknowledges financial support from the DFG SFB TR 103 through project A4. BG acknowledges financial support from the ERC-CoG-SHINE-771602. Authors also acknowledge use of the ZGH infrastructure (FIB Thermo Fisher Scientific, Helios G4 CX).

[1] M. Kolb, L.P. Freund, F. Fischer, I. Povstugar, S.K. Makineni, B. Gault, D. Raabe, J. Müller, E. Spiecker, S. Neumeier, M. Göken, On the grain boundary strengthening effect of boron in γ/γ' Cobalt-base superalloys, Acta Mater. 145 (2018) 247–254. https://doi.org/10.1016/j.actamat.2017.12.020.

[2] P. Kontis, H.A.M. Yusof, S. Pedrazzini, M. Danaie, K.L. Moore, P.A.J. Bagot, M.P. Moody, C.R.M. Grovenor, R.C. Reed, On the effect of boron on grain boundary character in a new polycrystalline superalloy, Acta Mater. 103 (2016) 688–699. https://doi.org/10.1016/j.actamat.2015.10.006.

[3] D. Tytko, P.-P. Choi, J. Klöwer, A. Kostka, G. Inden, D. Raabe, Microstructural evolution of a Ni-based superalloy (617B) at 700°C studied by electron microscopy and atom probe tomography, Acta Mater. 60 (2012) 1731–1740. https://doi.org/10.1016/j.actamat.2011.11.020.

[4] T. Alam, P.J. Felfer, M. Chaturvedi, L.T. Stephenson, M.R. Kilburn, J.M. Cairney, Segregation of B, P, and C in the Ni-Based Superalloy, Inconel 718, Metall. Mater. Trans. A. 43 (2012) 2183–2191. https://doi.org/10.1007/s11661-012-1085-9.

[5] X.B. Hu, H.Y. Niu, X.L. Ma, A.R. Oganov, C.A.J. Fisher, N.C. Sheng, J.D. Liu, T. Jin, X.F. Sun, J.F. Liu, Y. Ikuhara, Atomic-scale observation and analysis of chemical ordering in M3B2 and M5B3 borides, Acta Mater. 149 (2018) 274–284. https://doi.org/10.1016/j.actamat.2018.02.055.

[6]  P.A.J. Bagot, O.B.W. Silk, J.O. Douglas, S. Pedrazzini, D.J. Crudden, T.L. Martin, M.C. Hardy, M.P. Moody, R.C. Reed, An Atom Probe Tomography study of site preference and partitioning in a nickel-based superalloy, Acta Mater. 125 (2017) 156–165. https://doi.org/10.1016/j.actamat.2016.11.053.

[7]  M. Thuvander, K. Stiller, Microstructure of a boron containing high purity nickel-based alloy 690, Mater. Sci. Eng. A. 281 (2000) 96–103. https://doi.org/10.1016/S0921-5093(99)00741-8.

[8]  X.B. Hu, Y.L. Zhu, X.L. Ma, Crystallographic account of nano-scaled intergrowth of $M_2B$-type borides in nickel-based superalloys, Acta Mater. 68 (2014) 70–81. https://doi.org/10.1016/j.actamat.2014.01.002.

[9]  H.R. Zhang, O.A. Ojo, M.C. Chaturvedi, Nanosize boride particles in heat-treated nickel base superalloys, Scr. Mater. 58 (2008) 167–170. https://doi.org/10.1016/j.scriptamat.2007.09.049.

[10] M.J. Kaufman, V.I. Levit, Characterization of chromium boride precipitates in the commercial superalloy GTD 111 after long-term exposure, Philos. Mag. Lett. 88 (2008) 259–267. https://doi.org/10.1080/09500830801905445.

[11] T.J. Garosshen, T.D. Tillman, G.P. McCarthy, Effects of B, C, and Zr on the structure and properties of a P/M nickel base superalloy, Metall. Trans. A. 18 (1987) 69–77. https://doi.org/10.1007/BF02646223.

[12] L. Xiao, M.C. Chaturvedi, D. Chen, Effect of boron and carbon on the fracture toughness of IN 718 superalloy at room temperature and 650 °C, J. Mater. Eng. Perform. 14 (2005) 528. https://doi.org/10.1361/105994905X56106.

[13] B.C. Yan, J. Zhang, L.H. Lou, Effect of boron additions on the microstructure and transverse properties of a directionally solidified superalloy, Mater. Sci. Eng. A. 474 (2008) 39–47. https://doi.org/10.1016/j.msea.2007.05.082.

[14] P. Kontis, E. Alabort, D. Barba, D.M. Collins, A.J. Wilkinson, R.C. Reed, On the role of boron on improving ductility in a new polycrystalline superalloy, Acta Mater. 124 (2017) 489–500. https://doi.org/10.1016/j.actamat.2016.11.009.

[15] P. Kontis, A. Kostka, D. Raabe, B. Gault, Influence of composition and precipitation evolution on damage at grain boundaries in a crept polycrystalline Ni-based superalloy, Acta Mater. 166 (2019) 158–167. https://doi.org/10.1016/j.actamat.2018.12.039.

[16] H.L. Ge, J.D. Liu, S.J. Zheng, Y.T. Zhou, Q.Q. Jin, X.H. Shao, B. Zhang, Y.Z. Zhou, X.L. Ma, Boride-induced dislocation channeling in a single crystal Ni-based superalloy, Mater. Lett. 235 (2019) 232–235. https://doi.org/10.1016/j.matlet.2018.10.039.

[17] R.R. Unocic, N. Zhou, L. Kovarik, C. Shen, Y. Wang, M.J. Mills, Dislocation decorrelation and relationship to deformation microtwins during creep of a γ' precipitate strengthened Ni-based superalloy, Acta Mater. 59 (2011) 7325–7339. https://doi.org/10.1016/j.actamat.2011.07.069.

[18] K. Shinagawa, T. Omori, K. Oikawa, R. Kainuma, K. Ishida, Ductility enhancement by boron addition in Co–Al–W high-temperature alloys, Scr. Mater. 61 (2009) 612–615. https://doi.org/10.1016/j.scriptamat.2009.05.037.

[19] E. Cadel, D. Lemarchand, S. Chambreland, D. Blavette, Atom probe tomography investigation of the microstructure of superalloys N18, Acta Mater. 50 (2002) 957–966. https://doi.org/10.1016/S1359-6454(01)00395-0.

[20] Y.S. Zhao, J. Zhang, Y.S. Luo, B. Zhang, G. Sha, L.F. Li, D.Z. Tang, Q. Feng, Improvement of grain boundary tolerance by minor additions of Hf and B in a second generation single crystal superalloy, Acta Mater. 176 (2019) 109–122. https://doi.org/10.1016/j.actamat.2019.06.054.

[21] X. Wu, S.K. Makineni, P. Kontis, G. Dehm, D. Raabe, B. Gault, G. Eggeler, On the segregation of Re at dislocations in the γ' phase of Ni-based single crystal superalloys, Materialia. 4 (2018) 109–114. https://doi.org/10.1016/j.mtla.2018.09.018.

[22] S.K. Makineni, M. Lenz, S. Neumeier, E. Spiecker, D. Raabe, B. Gault, Elemental segregation to antiphase boundaries in a crept CoNi-based single crystal superalloy, Scr. Mater. 157 (2018) 62–66. https://doi.org/10.1016/j.scriptamat.2018.07.042.

[23] S.K. Makineni, A. Kumar, M. Lenz, P. Kontis, T. Meiners, C. Zenk, S. Zaefferer, G. Eggeler, S. Neumeier, E. Spiecker, D. Raabe, B. Gault, On the diffusive phase transformation mechanism assisted by extended dislocations during creep of a single crystal CoNi-based superalloy, Acta Mater. 155 (2018) 362–371. https://doi.org/10.1016/j.actamat.2018.05.074.

[24] P. Kontis, Z. Li, D.M. Collins, J. Cormier, D. Raabe, B. Gault, The effect of chromium and cobalt segregation at dislocations on nickel-based superalloys, Scr. Mater. 145 (2018) 76–80. https://doi.org/10.1016/j.scriptamat.2017.10.005.

[25] P. Kontis, D.M. Collins, A.J. Wilkinson, R.C. Reed, D. Raabe, B. Gault, Microstructural degradation of polycrystalline superalloys from oxidized carbides and implications on crack initiation, Scr. Mater. 147 (2018) 59–63. https://doi.org/10.1016/j.scriptamat.2017.12.028.

[26] Y.M. Eggeler, J. Müller, M.S. Titus, A. Suzuki, T.M. Pollock, E. Spiecker, Planar defect formation in the γ' phase during high temperature creep in single crystal CoNi-base superalloys, Acta Mater. 113 (2016) 335–349. https://doi.org/10.1016/j.actamat.2016.03.077.

[27] M.S. Titus, R.K. Rhein, P.B. Wells, P.C. Dodge, G.B. Viswanathan, M.J. Mills, A. Van der Ven, T.M. Pollock, Solute segregation and deviation from bulk thermodynamics at nanoscale crystalline defects, Sci. Adv. 2 (2016). https://doi.org/10.1126/sciadv.1601796.

[28] T.M. Smith, Y. Rao, Y. Wang, M. Ghazisaeidi, M.J. Mills, Diffusion processes during creep at intermediate temperatures in a Ni-based superalloy, Acta Mater. 141 (2017) 261–272. https://doi.org/10.1016/j.actamat.2017.09.027.

[29] S. Lu, S. Antonov, L. Li, C. Liu, X. Zhang, Y. Zheng, H.L. Fraser, Q. Feng, Atomic structure and elemental segregation behavior of creep defects in a Co-Al-W-based single crystal superalloys under high temperature and low stress, Acta Mater. 190 (2020) 16–28. https://doi.org/10.1016/j.actamat.2020.03.015.

[30] D. Barba, S. Pedrazzini, A. Vilalta-Clemente, A.J. Wilkinson, M.P. Moody, P.A.J. Bagot, A. Jérusalem, R.C. Reed, On the composition of microtwins in a single crystal nickel-based superalloy, Scr. Mater. 127 (2017) 37–40. https://doi.org/10.1016/j.scriptamat.2016.08.029.

[31] G.B. Viswanathan, R. Shi, A. Genc, V.A. Vorontsov, L. Kovarik, C.M.F. Rae, M.J. Mills, Segregation at stacking faults within the γ' phase of two Ni-base superalloys following intermediate temperature creep, Scr. Mater. 94 (2015) 5–8. https://doi.org/10.1016/j.scriptamat.2014.06.032.

[32] P. Kontis, Z. Li, M. Segersäll, J.J. Moverare, R.C. Reed, D. Raabe, B. Gault, The Role of Oxidized Carbides on Thermal-Mechanical Performance of Polycrystalline Superalloys, Metall. Mater. Trans. A. 49 (2018) 4236–4245. https://doi.org/10.1007/s11661-018-4709-x.

[33] S. Hamadi, F. Hamon, J. Delautre, J. Cormier, P. Villechaise, S. Utada, P. Kontis, N. Bozzolo, Consequences of a Room-Temperature Plastic Deformation During Processing on Creep Durability of a Ni-Based SX Superalloy, Metall. Mater. Trans. A. 49 (2018) 4246–4261. https://doi.org/10.1007/s11661-018-4748-3.

[34] X. Wu, S.K. Makineni, C.H. Liebscher, G. Dehm, J. Rezaei Mianroodi, P. Shanthraj, B. Svendsen, D. Bürger, G. Eggeler, D. Raabe, B. Gault, Unveiling the Re effect in Ni-based single

crystal superalloys, Nat. Commun. 11 (2020) 389. https://doi.org/10.1038/s41467-019-14062-9.

[35] A. Cervellon, S. Hémery, P. Kürnsteiner, B. Gault, P. Kontis, J. Cormier, Crack initiation mechanisms during very high cycle fatigue of Ni-based single crystal superalloys at high temperature, Acta Mater. 188 (2020) 131–144. https://doi.org/10.1016/j.actamat.2020.02.012.

[36] K. Roar, MacTempas, n.d. http://www.totalresolution.com.

[37] K. Thompson, D. Lawrence, D.J. Larson, J.D. Olson, T.F. Kelly, B. Gorman, In situ site-specific specimen preparation for atom probe tomography, Ultramicroscopy. 107 (2007) 131–139. https://doi.org/10.1016/j.ultramic.2006.06.008.

[38] R.C. Reed, A.C. Yeh, S. Tin, S.S. Babu, M.K. Miller, Identification of the partitioning characteristics of ruthenium in single crystal superalloys using atom probe tomography, Scr. Mater. 51 (2004) 327–331. https://doi.org/10.1016/j.scriptamat.2004.04.019.

[39] I. Goldfarb, W.D. Kaplan, S. Ariely, M. Bamberger, Fault-induced polytypism in (Cr, Fe)2B, Philos. Mag. A. 72 (1995) 963–979. https://doi.org/10.1080/01418619508239947.

[40] X.B. Hu, Y.L. Zhu, X.H. Shao, H.Y. Niu, L.Z. Zhou, X.L. Ma, Atomic configurations of various kinds of structural intergrowth in the polytypic M2B-type boride precipitated in the Ni-based superalloy, Acta Mater. 100 (2015) 64–72. https://doi.org/10.1016/j.actamat.2015.08.025.

[41] A.G. Fitzgerald, M. Mannami, Electron Diffraction From Crystal Defects: Fraunhofer Effects From Plane Faults, Proc. R. Soc. Lond. Ser. Math. Phys. Sci. 293 (1966) 169–180.

[42] M.J. Whelan, P.B. Hirsch, Electron diffraction from crystals containing stacking faults: I, Philos. Mag. J. Theor. Exp. Appl. Phys. 2 (1957) 1121–1142. https://doi.org/10.1080/14786435708242742.

[43] F. Meisenkothen, E.B. Steel, T.J. Prosa, K.T. Henry, R. [Prakash Kolli, Effects of detector dead-time on quantitative analyses involving boron and multi-hit detection events in atom probe tomography, Ultramicroscopy. 159 (2015) 101–111. https://doi.org/10.1016/j.ultramic.2015.07.009.

[44] B. Gault, D.W. Saxey, M.W. Ashton, S.B. Sinnott, A.N. Chiaramonti, M.P. Moody, D.K. Schreiber, Behavior of molecules and molecular ions near a field emitter, New J. Phys. 18 (2016) 033031. https://doi.org/10.1088/1367-2630/18/3/033031.

[45] Z. Peng, F. Vurpillot, P.-P. Choi, Y. Li, D. Raabe, B. Gault, On the detection of multiple events in atom probe tomography, Ultramicroscopy. 189 (2018) 54–60. https://doi.org/10.1016/j.ultramic.2018.03.018.

[46] E.A. Marquis, F. Vurpillot, Chromatic Aberrations in the Field Evaporation Behavior of Small Precipitates, Microsc. Microanal. 14 (2008) 561–570. https://doi.org/10.1017/S1431927608080793.

[47] O.C. Hellman, J.A. Vandenbroucke, J. Rüsing, D. Isheim, D.N. Seidman, Analysis of Three-dimensional Atom-probe Data by the Proximity Histogram, Microsc. Microanal. 6 (2000) 437–444. https://doi.org/10.1007/S100050010051.

[48] B. Jiang, C. Chen, X. Wang, H. Wang, W. Wang, H. Ye, K. Du, Deformation induced twinning and phase transition in an interstitial intermetallic compound niobium boride, Acta Mater. 165 (2019) 459–470. https://doi.org/10.1016/j.actamat.2018.12.011.

[49] X.B. Hu, Y.L. Zhu, N.C. Sheng, X.L. Ma, The Wyckoff positional order and polyhedral intergrowth in the M3B2- and M5B3-type boride precipitated in the Ni-based superalloys, Sci. Rep. 4 (2014) 7367. https://doi.org/10.1038/srep07367.

[50] M.C. Paulisch, N. Wanderka, G. Miehe, D. Mukherji, J. Rösler, J. Banhart, Characterization of borides in Co–Re–Cr-based high-temperature alloys, J. Alloys Compd. 569 (2013) 82–87. https://doi.org/10.1016/j.jallcom.2013.03.086.

[51] A.K. da Silva, G. Leyson, M. Kuzmina, D. Ponge, M. Herbig, S. Sandlöbes, B. Gault, J. Neugebauer, D. Raabe, Confined chemical and structural states at dislocations in Fe-9wt%Mn steels: A correlative TEM-atom probe study combined with multiscale modelling, Acta Mater. 124 (2017) 305–315. https://doi.org/10.1016/j.actamat.2016.11.013.

[52] M. Kuzmina, M. Herbig, D. Ponge, S. Sandlöbes, D. Raabe, Linear complexions: Confined chemical and structural states at dislocations, Science. 349 (2015) 1080–1083. https://doi.org/10.1126/science.aab2633.

[53] S. Lartigue-Korinek, M. Walls, N. Haneche, L. Cha, L. Mazerolles, F. Bonnet, Interfaces and defects in a successfully hot-rolled steel-based composite Fe–TiB2, Acta Mater. 98 (2015) 297–305. https://doi.org/10.1016/j.actamat.2015.07.024.

[54] W.D. Summers, E. Alabort, P. Kontis, F. Hofmann, R.C. Reed, In-situ high-temperature tensile testing of a polycrystalline nickel-based superalloy, Mater. High Temp. 33 (2016) 338–345. https://doi.org/10.1080/09603409.2016.1180857.

[55] X.Z. Qin, J.T. Guo, C. Yuan, C.L. Chen, H.Q. Ye, Effects of Long-Term Thermal Exposure on the Microstructure and Properties of a Cast Ni-Base Superalloy, Metall. Mater. Trans. A. 38 (2007) 3014–3022. https://doi.org/10.1007/s11661-007-9381-5.